\documentclass{aa}  
\usepackage{graphicx}
\usepackage{aalongtable}
\usepackage{natbib}

\newcommand{\belines}{\ion{Be}{ii} UV doublet lines at 313.0-1 nm}
\newcommand{\teff}{$T_\mathrm{eff}$}
\newcommand{\cdthir}{\object{$\mathrm{CD}-30\degr\,0298$}}

\newcommand{\oi}{O\,{\sc i}}
\newcommand{\liline}{Li line at 670.78 nm}
\bibpunct{(}{)}{;}{a}{}{,}
\begin{document}
   \title{Li and Be depletion in metal-poor subgiants\thanks{Based on observations collected at the European Southern 
Observatory, Chile (ESO No. 68.D-0546)}}
   \subtitle{}
 
   \author{A.E. Garc\'{\i}a P\'erez
          \inst{1}{$^,$\thanks{Student visitor at the European Southern Observatory, 
Munich, Germany}}
          \and
          F. Primas\inst{2}}

   \offprints{A.E.Garcia-Perez@open.ac.uk}
 
   \institute{Department of Astronomy and Space Physics, Uppsala University,
              Box 515, 751\,20 Uppsala, Sweden\\
              \email{aegp@astro.uu.se}
         \and
          European Southern Observatory, Karl-Schwarzschild Strasse 2, 85748 
Garching bei M\"unchen, Germany  \\
             \email{fprimas@eso.org}
                          }
   
\date{Received soon; accepted quite soon}
%\titlerunning{Li and Be depletion in metal-poor subgiants}
\authorrunning{A.E. Garc\'{\i}a P\'erez and F. Primas}
   \abstract{A sample of metal-poor subgiants has been
observed with the UVES spectrograph at the Very Large Telescope 
and abundances of Li and Be have been determined. Typical
signal-to-noise per spectral bin values for the co-added spectra are
of the order of 500 for the \ion{Li}{i} line (670.78 nm) and
100 for the \ion{Be}{ii} doublet lines (313.04 nm). The spectral
analysis of the observations was carried out using the Uppsala suite of
 codes and {\sc{marcs}} (1D-LTE) model atmospheres with stellar parameters
from photometry, parallaxes, isochrones and Fe~{\sc{ii}} lines. Abundance 
estimates of the light elements were corrected for departures from
local thermodynamic equilibrium in the line formation. Effective
temperatures and Li abundances  seem to be correlated and Be
abundances correlate with [O/H]. Standard models predict Li and Be
abundances approximately one order of magnitude lower than
main-sequence values which is in general agreement with the
observations. On average, our observed depletions seem to be 0.1 dex
smaller and between 0.2 and 0.4 dex larger (depending on which
reference is taken) than those predicted for Li and Be, respectively.
This is not surprising since the initial Li abundance, as derived from
main-sequence stars on the Spite plateau, may be systematically in
error by 0.1 dex or more, and uncertainties in the spectrum 
normalisation and continuum drawing may affect our Be abundances
systematically.

   }
   
   \maketitle
%
%________________________________________________________________

\section{Introduction}

Stellar abundances of the very light elements lithium, beryllium, and
boron (Li, Be, and B) have been studied extensively during the recent
decades, mainly because of the great role they play in constraining Big 
Bang nu\-cleo\-syn\-the\-sis theories as well as models of galactic chemical 
evolution and stellar evolution. 

Stellar evolution predicts depletion of the light elements in stellar 
atmospheres under certain physical conditions. Stars with their surface 
convection base reaching depths where there has been some
depletion should reflect such depletion in their stellar surfaces. The 
convective motions can bring up material from the interior to the surface,
i.e. gas poor in light elements is mixed with gas with the initial 
composition, leading to dilution, this is expected to happen in subgiant stars 
which are the type of stars under study in this work. If the 
temperature at the base of the convection zone is high enough, light
elements brought down from the surface can be destroyed.

With this background in mind, we have looked at lithium and beryllium
abundances in a sample of metal-poor subgiant stars, from a new set
of UVES observations. The analysis of the Be features, in crowded 
spectral regions where blending may be a serious problem, requires 
detailed spectrum synthesis of high quality. 

Standard models, e.g. \citet{Deliyannis90}, include only convection as
the mixing mechanism. According to these models, depletion occurs at
low stellar mass and low effective temperatures. These stars have
convection zones deep enough to reach inner layers where the light
elements have been depleted. The models predict burning of the 
light isotopes during the pre-main sequence (PMS) phase and dilution 
in the subgiant evolutionary phase. The fact that the Sun has a
photospheric logarithmic Li abundance of $A($Li$)=1.05$
\citep{Asplundb05} which is significantly lower than the values 
measured in meteoritic matter (which is thought to be representative
of the gas at the time our solar system formed) indicates that also 
main-sequence (MS) stars undergo some mixing process (other
than simple convection). 

On the main-sequence, diffusion \citep{Michaud86,Richard02}, rotation induced 
mixing \citep{Pinsonneault90}, mass loss \citep{Schramm90,Vauclair95}
and gravity waves \citep{ramon91,Talon04} have been proposed to explain
the light element patterns observed in MS stars. Light elements 
burn at low, but different, burning temperatures. The most fragile of them is 
Li (with a burning temperature $T_{\rm b}\simeq2.5\times10^6~K$), followed by 
Be ($T_{\rm b}\simeq3.0\times10^6~K$) and B ($T_{\rm  b}\simeq5.0\times10^6~K$).
Therefore, if at least the abundances of two of the light isotopes are known 
in a star, their abundances may be used to further constrain which type of 
internal mixing is active in the stellar external layers. Models with 
only mass loss fail to predict a simultaneous Li and Be depletion, i.e. 
lithium is expected to be completely depleted before beryllium shows any sign 
of depletion. However observations show just the opposite, i.e. some degree
of Be depletion 
in stars in which Li can still be detected. One such example (there are 
several other cases) are the Hyades F stars, which show depletion in both Li 
and Be \citep{Boesgaard86a,Deliyannis98}. 
 
As for the case of  main-sequence stars, the knowledge of two or more light 
isotopes in subgiants can help us to better understand the mixing processes 
occurring in the atmospheres of these stars. In this work, we have exploited 
that possibility by determining both Li and Be abundances and comparing 
the observed abundance depletions with predictions based on standard models. 

This comparison requires the assumption of a value for the initial 
abundances so that the magnitude of the observed depletion can be quantified. 
About $70\%$ of the \element[][7]{Li} observed in the early Galaxy is 
synthesised during the Big Bang \citep{Walker91,Walker93,Olive00}. Some extra 
contributions by $\alpha-\alpha$ reactions during late stellar 
evolutionary stages, like (post-)AGB stars and novae, contribute at later 
epochs to the enrichment of the Galaxy. These sources influence the galactic
evolution of lithium at higher metallicities. Therefore, \element[][7]{Li}
abundances in the oldest, most metal-poor stars of the Galactic halo are
of great importance from a cosmological point of view, providing useful
constraints on the cosmological baryon-to-photon ratio $\eta$. In the early
1980's, \citet{Spite82} found a constant logarithmic lithium abundance, 
$A(\mathrm{Li})=2.1$ \footnote{Abundances are denoted by 
$A(\mathrm{X})=\log{(N_{\rm X} / N_{\rm H})}+12$}, in
a small sample of warm metal-poor main-sequence stars spanning a wide
range of (low) metallicities, the so called Spite lithium plateau. Although
the plateau has now been confirmed by large data samples, the interpretation
of its constant value (whether Li is primordial or not) still remains an
open question, as well as if it is tilted with respect to metallicity and
effective temperature \citep[cf][for a recent review]{Charbonnel05}. If
this plateau value represents the primordial abundance of Li, then one
needs to explain why in the Galaxy today its abundance is 3.25
\citep{Asplundb05}. 

For the other light elements, \citet{Reeves70} and \citet{Meneguzzi71} 
proposed cosmic-ray spallation reactions as the main (if not unique, as
in the case of beryllium) source of \element[][6]{Li}, \element[][9]{Be}
and \element[][10,11]{B} 
production. In this scenario, light elements are produced via collisions
of C and O target nuclei with protons and $\alpha$ particles in the
interstellar medium. This predicts a quadratic dependence of the light
isotope abundance on the target nuclei, e.g. oxygen. However, in the early
1990's, \citet{Gilmore92} (for Be) and \citet{Duncan92} (for B) found
a linear slope, which has triggered the exploration of several variations of
the {\it standard} cosmic-ray spallation theory. The observed linear trend 
(further confirmed by larger data samples, \citet{Molaro97b}, \citet{BoesBe99})
seem to imply a primary origin of Be and B instead of a secondary one. One
possibility to achieve this is if C and O nuclei were freshly synthesised
by Type II supernovae (SNeII) and accelerated by the outwards shocks together
with the $\alpha$ particles and protons, and different scenarios more or less
built on this idea have been exploited
\citep{Duncan92,Feltzing94,Casse95,Parizot99}. Knowledge of the correlation
between Be and O stellar abundances in larger samples of stars will help us to
discern between these different scenarios (it should be noted that many
more models have been developed in the last few years, and that those mentioned
above represent some main ideas, which have subsequently been developed
with different degrees of sophistication).

On the observational side, in general, the 
abundances of light elements in stars are so small 
that the only lines used for abundance determination purposes are resonance 
lines. The observations of these lines as well as their modelling have been 
quite a challenge. Most studies have so far concentrated on lithium; 
its resonant doublet at 670.78 nm makes it easily accessible with ground 
based telescopes. Beryllium, 
instead, having its main resonant doublet at 313.0 nm, requires high
resolution and highly efficient spectrographs in the near-UV. Boron,
the third light element, is even more challenging as it may be studied
only from space (such as with the Hubble Space Telescope), as 
its main resonant doublet falls at 250.0 nm. Thanks to the very efficient 
high-resolution spectrographs currently available, we note that observations 
of \element[][6]{Li} in Galactic stars have become more accurate and 
affordable in terms of telescope time \citep[] [and references therein]
{Nissen00,Aoki04}.

In the following sections we describe the observations as well as the  
stellar parameters and stellar abundances determinations, which are then
discussed in the last section in terms of dilution in subgiants. 

%__________________________________________________________________

\section{Observations and data reduction}

Our sample of metal-poor subgiants was observed in service mode between
October 2000 and March 2001, with the Ultraviolet Visual Echelle Spectrograph
\citep[UVES,][]{Dekker00}, mounted on the Nasmyth B platform of the 8\,m Kueyen
unit of the ESO Very Large Telescope (VLT). All observations were taken in
dichroic mode, i.e. using both the red and the blue arms of the spectrograph
simultaneously, thus covering a wide range in wavelength. The
instrument set-ups chosen were the Dichroic\#1 (Blue346 combined with Red580,
where 346 and 580 denote the central wavelengths in nm of the two
settings), and Dichroic \#2 (Blue346 plus Red860). In this way we obtained a
spectral coverage from 300 to 380 nm in the blue, and from 480 to 1000
nm in the red (with the only exception of the narrow gaps between the
two CCDs used in the UVES Red Arm). 

The observations had been proposed and planned for a detailed study of
oxygen abundances using three different abundance indicators 
\citep[{[O{\sc{i}}]}, O{\sc{i}}, and the OH lines in the near-UV,][]{aegpO}, but
the large spectral coverage also allowed us to study other elements
like the light elements, 
lithium and beryllium. The need for detecting the three different oxygen
indicators in the same object restricted this study to a relatively narrow range 
of surface gravities and effective temperatures (see Table~\ref{gripar}), which
corresponds to stars that have already left the MS and started evolving (ascending)
along the subgiant (giant) branch. Moreover, accurate oxygen abundances determinations require
accurate surface gravity determinations. Hence, our stars were chosen based on
the availability of Hipparcos parallaxes from which surface gravities were estimated.
The selected stars ended up being not very distant objects (the parallaxes of most
of them indicate distances smaller than $400\mathrm{kpc}$) which reduces the effective
temperature uncertainties due to reddening.

Lithium is easily measurable from its resonant doublet at
670.78 nm, whereas beryllium is more challenging as its main resonant doublet
falls at 313.0 nm, i.e. in a very crowded region and very close to the atmospheric
cut-off. In order to maximise the signal-to-noise ratio of the near-UV spectra, a 2$\times$2
binning was selected when reading out the EEV chip of the UVES blue arm. This choice
did not affect the resolution of our final spectra since we chose a slit width of
1{\arcsec}, the nominal resolving power of which ($R=45\,000$) implies some oversampling.
In the red arm, we selected a slit width of 0.7\arcsec corresponding to a nominal
resolving power of $R=55\,000$, thus resolving the blending of the Li line by an \ion{Fe}{i}
line at 670.74~nm (in spectra for which such blending could be a problem, i.e. for the more metal-rich 
stars). Typical signal-to-noise ($S/N$) per spectral bin values are of the order of 100 and 500 at the Be
and Li lines, respectively, in the final co-added spectra.

Data reduction followed standard reduction procedures, i.e. bias subtraction, division
by a normalised flat-field, extraction of the echelle orders and their wavelength
calibration. After a careful inspection of the blue spectra, we decided that the quality
achieved by the UVES data reduction pipeline processing was satisfactory for our 
purposes.
However,
for the red spectra, we decided to manually re-reduce all the spectra (using IRAF) in order to improve (i.e. minimise) the presence of fringes and ripples
in these very high $S/N$ spectra. We then normalised all the spectra with a spline
function. This step is straightforward in the lithium region (where the stellar continuum
is easily identifiable), but the severe spectral-line crowding in the near-UV region
does not allow such an easy identification. Instead, the near-UV spectra were normalised
by using synthetic spectra that were computed for each individual star. 

\begin{figure}
\begin{center}
\resizebox{\hsize}{!}
{\includegraphics{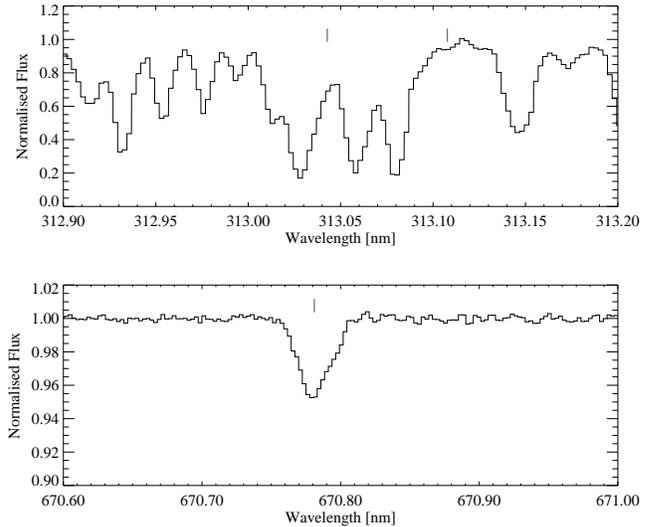}}
\caption{\label{obser} Spectra of the Be (top) and Li (bottom) regions in one star of
our sample, \object{HD\,218857}. The positions of the \ion{Be}{ii} doublet and of the \ion{Li}{i}
line are indicated by vertical bars in both panels.}
\end{center}
\end{figure}

In spite of the weakness of the observed lines, the very high quality of our UVES spectra
allowed us to detect both the Li and the Be lines in almost all stars. Fig.~\ref{obser} 
illustrates the spectral quality of one of our targets, \object{HD\,218857} ($V=8.967$).

%______________________________

\section{Stellar parameters and model atmospheres}

%______________________________

\subsection{Stellar parameters}

Since the determination of the stellar parameters for our sample of stars 
is presented in a parallel paper \citep{aegpO}, here we will only summarise how they 
were derived. Table~\ref{gripar} lists the values used for the effective temperature ($T_\mathrm{eff}$),
the logarithmic surface gravity ($\log{g}$), the metallicity ([Fe/H]) and the microturbulence.

The process of determining stellar parameters is clearly an iterative process.
In short, $T_\mathrm{eff}$ was estimated from the $(b-y)$- and $(V-K)$-$T_\mathrm{eff}$
calibrations of \citet{Alonso96a}, based on the Infrared Flux Method (IRFM).
Str\"omgren ($uvby\beta$) and $V$, $K$ photometry were taken from Schuster \& Nissen
(1988, and private communication) \nocite{Schuster88})
 and the Two Micron All Sky Survey (2MASS) database, respectively. Colours and magnitudes were de-reddened
with values estimated from the \citet{Hakkila97} galactic maps of interstellar
reddening. 

The surface gravity was derived from Hipparcos parallaxes following the prescription 
in \citet{Nissen97}. Parallaxes are available for all programme stars and, furthermore,
their uncertainties are all very small, except for two stars, \object{HD\,126587} and 
\cdthir, for
which the uncertainty is as large or even larger than the parallax itself. Therefore, the 
surface gravities
of these two stars were estimated from a colour-magnitude diagram using \citet{VandenBerg00}
evolutionary tracks. As described in \citet{aegpO}, this method was applied to the
entire sample, as a consistency check on the $\log g$ values derived from the parallaxes.
A mass of $0.8\,M_{\sun}$ was assumed for all the stars and the bolometric corrections were 
taken from \citet{Bessell98}. 

The metallicity values reported in Table~\ref{gripar} are spectroscopic values, and they
have been determined from the equivalent widths of a set of \ion{Fe}{ii} lines 
\citep[for more details see][]{aegpO}. This same set of iron lines was used to constrain
the microturbulence, by minimising the dependence of the iron abundance on the
equivalent widths of the lines. 

\subsection{Model atmospheres}

In order to model the stellar atmospheres of the programme stars, the equation of
hydrostatic equilibrium, the equation of radiative transfer and the flux-constancy
condition, with mixing-length convective flux, were solved simultaneously for 
the set of stellar parameters of each of the stars (cf Table~\ref{gripar}), 
using an updated version of the {\sc{marcs}} code \citep{Asplund97}.
In order to solve the equations (in one dimension and assuming local thermodynamic equilibrium (LTE) conditions), the
program requires in addition to the stellar parameters listed above also 
abundances of certain elements such as C, N, O, Ne, Na, Mg, etc. 
These elements need to be taken into account when computing the number of electrons and
opacities in stellar atmospheres and their relative abundances were chosen to be solar 
\citep{Grevesse98}, scaled to the stellar metallicity; however, the solar oxygen 
abundance was taken from the 1D-LTE analysis of the oxygen 
forbidden line at 630.03 nm \citep{Nissen02}, and a value of 0.4~dex was assumed for the
$\alpha$-enhancement of Population II stars. No enhancement was applied to the C abundance.

\begin{table}
\caption{\label{gripar} Stellar parameters of our programme stars as derived by
\citet{aegpO}}
\centering
\begin{tabular}{lcccccc}
\hline\hline 
Star& $T_{\mathrm{eff}}$ & $\log{g}$ & [Fe/H] &$\xi_\mathrm{micro}$\\ 
&[K]&[cgs]&[dex]&[km.s$^{-1}$]\\
\hline
\object{HD\,4306}   & 4990 & 3.04$\pm$0.26 & $-2.33$ & 1.5 \\
\object{HD\,26169}  & 4972 & 2.49$\pm$0.26 & $-2.28$ & 1.5 \\
\object{HD\,27928}  & 5044 & 2.67$\pm$0.43 & $-2.14$ & 1.5 \\
\object{HD\,45282}  & 5352 & 3.15$\pm$0.13 & $-1.52$ & 1.1 \\
\object{HD\,108317} & 5300 & 2.76$\pm$0.22 & $-2.25$ & 1.5 \\
\object{HD\,126587} & 4712 & 1.66$\pm$0.50 & $-2.87$ & 1.5 \\
\object{HD\,128279} & 5336 & 2.95$\pm$0.21 & $-2.19$ & 1.5 \\
\object{HD\,200654} & 5292 & 2.86$\pm$0.35 & $-2.73$ & 1.5 \\
\object{HD\,218857} & 5015 & 2.78$\pm$0.36 & $-1.79$ & 1.4 \\
\object{HD\,274939} & 5090 & 2.79$\pm$0.33 & $-1.49$ & 1.3 \\
\object{$\mathrm{BD}-01\degr\,2582$} & 5072 & 2.92$\pm$0.40 & $-2.11$ & 1.5 \\
\object{$\mathrm{CD}-24\degr\,1782$} & 5228 & 3.46$\pm$0.35 & $-2.25$ & 1.5 \\
\object{$\mathrm{CD}-30\degr\,0298$} & 5196 & 2.93$\pm$0.50 & $-3.01$ & 1.5 \\
\hline
\end{tabular} 
\end{table}

\section{Abundance analysis and uncertainties}

\subsection{Lithium abundances}
The {\liline} lies in a spectral region which is almost 
free of other lines, making it possible
in all our spectra to easily identify a clean wavelength region of $\sim$ 0.4 nm
around the Li line and to unambiguously define the stellar continuum. This, together
with the fact that the blending by a neighbouring {\ion{Fe}{i}} line at 670.74 nm is 
insignificant at our low metallicities, makes the measurement of
the equivalent widths very accurate. The widths were measured by integrating the area under
the line using the {\sc{iraf}} task {\it{splot}} and assuming a Gaussian profile. 
Direct integration, without the profile assumption, gave very similar results. 
The values are given in Col.~2 of Table~\ref{Li}. There is only one star, 
\object{HD\,108317}, for which the line could not be detected. A value of $0.08$ pm 
was estimated as an upper limit to its equivalent width based on a two sigma 
observational error. The associated abundance must therefore be considered an upper limit.

The lithium line at 670.78 nm is a multiplet but the resolution of our observed red
spectra is not high enough to resolve it completely. The feature was treated as a single
line when solving the radiative transfer equation. The oscillator strength used
in our analysis ($\log gf = 0.17$) was the sum of the values for the individual 
components given in Table~3 of \citet{Smith93}. 1D-LTE abundances were derived
from the measured equivalent widths using the {\sc{eqw}} code from the Uppsala package
of stellar atmospheres. 
As a cross-check, we also ran some spectral syntheses and found that in
general there is a very good agreement between the lithium abundances derived
by the two methods.
However, it is well known that the \ion{Li}{i} line at 670.78 nm
forms under non-LTE conditions (NLTE). Therefore, NLTE corrections should be applied
to the LTE abundances. For the stars analysed here, these corrections were estimated
using the routines and the NLTE results in \citet{Carlsson94}. The final NLTE Li
abundances (i.e. LTE Li plus the correction) are listed in Col.~9 of Table~\ref{Li}. 
The largest NLTE correction is of the order of 0.20 dex. The lithium 
abundance of \object{HD\,108317} and the effective temperature of the giant \object{HD\,126587}
 lie outside the range of \teff\ and $A({\mathrm{Li}})$ covered in \citet{Carlsson94}. 
For these two stars, NLTE corrections based on their closest points in the 
grid of Carlsson et al. have been adopted.

\begin{table*}
\caption{\label{Li} Lithium abundances of the programme stars. Equivalent widths
($W_\lambda$) of the Li feature at 670.78 nm are given in the second column, followed
in the third column by the lithium abundances. Columns 4 to 7 show the sensitivity
of the lithium abundances to changes in the stellar parameters and in the equivalent
width (varied by their typical uncertainties). Final errors based on the square 
sum of these values are listed in Col.~8 ($\sigma$). NLTE Li abundances as derived
from the {\sc multi} code of \citet{Carlsson94} and the oxygen abundances as derived
from the {\oi} lines at 630.03 nm \citep[cf][]{aegpO} are given in the two last
columns.}
\begin{center}
\begin{tabular}{lcccccccccc} 
\hline\hline 
Star & $W_\lambda$ & $A(\mathrm{Li})$ & $\Delta{A(\mathrm{Li})}$ & $\Delta{A(\mathrm{Li})}$ & $\Delta{A(\mathrm{Li})}$ & $\Delta{A(\mathrm{Li})}$ & $\sigma$ &\ $A(\mathrm{Li})$& $A(\mathrm{O})$\\
 & [pm] & LTE & $(T_{\mathrm{eff}})$ & ($\log{g}$) & ([Fe/H]) & ($W_\lambda$) & & NLTE & LTE\\
\hline
\object{HD\,4306}     &  1.63   &    0.97  & 0.11 &  0.00 &  0.00 & 0.02 & 0.11 & 1.06 & 7.14 \\   
\object{HD\,26169}    &  1.44   &    0.91  & 0.11 & $-0.01$ &  0.00 & 0.02 & 0.11 & 1.08 & 7.05 \\   
\object{HD\,27928}    &  1.36   &    0.96  & 0.10 & $-0.02$ & $-0.01$ & 0.01 & 0.10 & 1.12 & 7.06 \\   
\object{HD\,45282}    &  1.03   &    1.14  & 0.10 &  0.00 &  0.00 & 0.02 & 0.10 & 1.24 & 7.72 \\   
\object{HD\,108317}   & $<0.08$ & $ <0.00$ & 0.09 & $-0.01$ &  0.00 & $<0.18$ & 0.20 & $<0.11$ & 7.23 \\  
\object{HD\,126587}   &  1.75   &    0.76  & 0.11 & $-0.03$ &  0.00 & 0.01 & 0.11 & 0.96 & 6.32 \\   
\object{HD\,128279}   &  1.06   &    1.13  & 0.09 &  0.00 &  0.00 & 0.02 & 0.09 & 1.24 & 6.96 \\   
\object{HD\,200654}   &  1.10   &    1.09  & 0.10 &  0.00 &  0.01 & 0.03 & 0.10 & 1.20 &$<6.75$\\  
\object{HD\,218857}   &  1.33   &    0.90  & 0.12 & $-0.01$ &  0.00 & 0.02 & 0.12 & 1.06 & 7.33 \\   
\object{HD\,274939}   &  1.44   &    1.02  & 0.11 & $-0.01$ &  0.00 & 0.02 & 0.11 & 1.17 & 7.76 \\   
\object{$\mathrm{BD}-01\degr\,2582$} & 0.63 & 0.63 & 0.10 & $-0.01$ & 0.00 & 0.03 & 0.10 & 0.77 &  $7.17$ \\
\object{$\mathrm{CD}-24\degr\,1782$} & 1.12 & 1.05 & 0.09 &  0.00 & 0.00 & 0.02 & 0.09 & 1.16 &  6.80 \\
%\footnote{There was not 
%oxygen abundance determined in \citet{aegpO} 
%from the forbidden line so the value based on OH UV 
%line was assumed instead} \\
\object{$\mathrm{CD}-30\degr\,0298$} & 1.01 & 0.96 & 0.09 &  0.00 & 0.00 & 0.02 & 0.09 & 1.08 & $<6.38$ \\
\hline
\end{tabular}
\end{center}
\end{table*}

The final uncertainty in each lithium abundance was calculated by taking into
account the uncertainty in the equivalent width measurement and in the stellar
parameters. According to photon statistics, the associated error in equivalent
widths for a $S/N \sim 500$ spectrum is of the order of $0.03-0.04$ pm. Here,
in order to take into account also the uncertainty due to the continuum placement
(although very small) we assumed a total uncertainty of $0.05$ pm in each
equivalent width. On the average, this corresponds to abundance errors of the
order of 0.02 dex only.

The sensitivity to the stellar parameters was computed by testing the
response of the derived lithium abundance to changes in $T_\mathrm{eff}$
($\pm$100 K), in $\log g$ ($\pm$ its 1$\sigma$ errors which are listed in Col.~3
of Table~\ref{gripar}), and in [Fe/H] ($\pm$ 0.1 dex). The influence of these
changes on the final Li abundance are listed in Cols.~4 to 7 of Table~\ref{Li}.
Obviously, the lithium abundances are mostly sensitive to
the effective temperature (a 0.1 dex effect). The errors in lithium abundances
due to errors in other stellar parameters are negligible.

\subsection{Beryllium abundances}

The beryllium abundances are based on synthetic-spectrum analysis of the \ion{Be}{ii}
resonant doublet at 313.0-1 nm. The bluer component ($\lambda$\,313.04 nm) is the
stronger of the two lines, but it is severely blended (mainly by Cr, V, CH and
OH lines). The other component of the doublet ($\lambda$\,313.11 nm) is very weak in
our programme stars; it lies in the wing of a much stronger Ti line but it is otherwise
almost free of blends. There is no special reason to prefer one line
over the other. Instead, spectral synthesis of a region of 0.3 nm
around the two Be lines is mandatory in order to properly account 
for the blends. The 1D-LTE syntheses were run with the
program {\sc{bsyn}} of the Uppsala suite of
codes. The line list of atomic lines and CH lines used was the same as the one in
\citet{Primas97}, whereas a new OH line list was used, based on \citet{Gillis01}.

\begin {figure*}
\begin{center}
\resizebox{\hsize}{!}
{\includegraphics{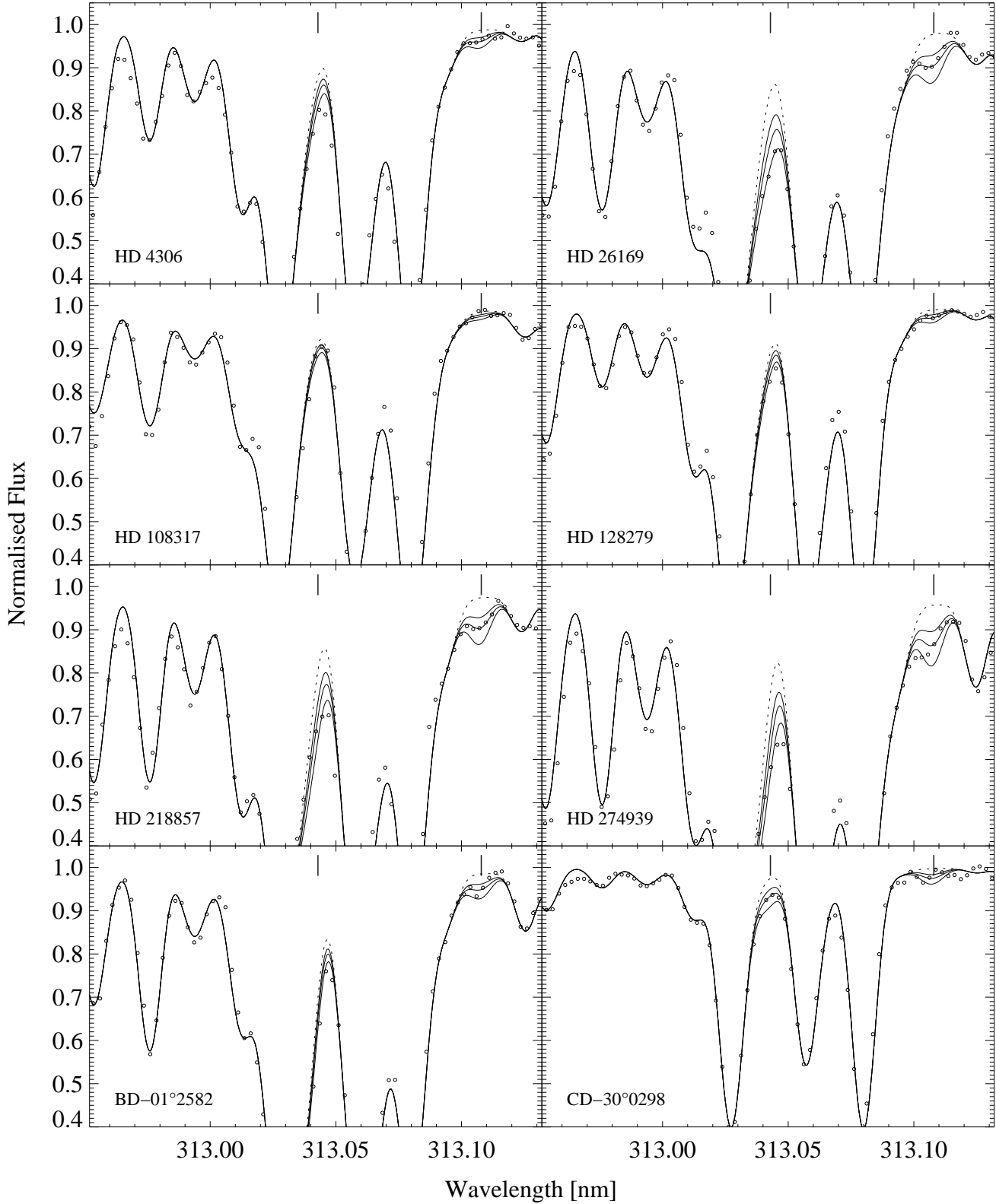}}
\caption{\label{syn} 1D-LTE spectral syntheses for a selection of 8 programme
stars, in the wavelength range around the {\belines} computed for four different
Be abundances: the best-fit abundance, abundance changes by $\pm0.2$ dex around that value, and a
very low value of the Be abundance (practically zero). The observed spectra are denoted
by open circles and the synthetic spectra by solid lines. The dashed line corresponds 
to the synthetic spectra for the very low value of $A(\mathrm{Be})$.}
\end{center}
\end{figure*}

The elemental abundances were those adopted for the model atmospheres, except
for those elements with transitions in the 0.3 nm wavelength range that was synthesised. 
Once computed, 
each synthetic spectrum was convolved with a profile taking care of macroturbulence, 
stellar rotation etc., and with a Gaussian with a width corresponding to the
resolution of the spectra. The widths of these functions were changed together with the 
stellar abundances until the best fit was achieved. Our final beryllium abundances 
(Col.~2, Table~\ref{Be}) correspond to the abundance value that best fits both 
\ion{Be}{ii} lines, see Fig.~\ref{syn}.

\begin{table}
\caption{\label{Be} Our final Be abundances as determined from the spectral synthesis
fitting of the Be doublet (Col.~2). The sensitivity of the result to changes in the stellar
parameters (corresponding to their uncertainties) are presented 
in Cols.~3 to 5 and their associated $\sigma$ values, 
calculated as in the case of Li, are given in Col.~6. Column~7 lists
the final NLTE corrected Be abundances.}
\begin{center}
 \renewcommand{\tabcolsep}{0.5pt}
\begin{tabular}{lcccccc}
\hline\hline 
Star & $A(\mathrm{Be})$ & $\Delta{A(\mathrm{Be})}$ & $\Delta{A(\mathrm{Be})}$ &
$\Delta{A(\mathrm{Be})}$ & $\sigma $& $A(\mathrm{Be})$ \\
 & LTE & $(T_{\mathrm{eff}})$ & ($\log{g}$) & ([Fe/H]) & &NLTE \\
\hline
\object{HD\,4306}   & $-1.84$  & $-0.02$ & 0.17 & 0.03 & 0.17& $-1.84$ \\                 
\object{HD\,26169}  & $-1.65$  & { 0.00} & 0.15 & 0.02 & 0.15& $-1.65$ \\                
\object{HD\,27928}  & $-1.96$  & $-0.01$ & 0.26 & 0.02 & 0.26 & $-2.00$ \\                
\object{HD\,45282}  & $-1.15$  & { 0.00} & 0.07 & 0.03 & 0.08 & $-1.18$ \\              
\object{HD\,108317} & $<-2.30$ & { 0.03} & 0.10 & 0.01 & 0.10 & $<-2.26$ \\           
\object{HD\,126587} & $-2.58$  & { 0.02} & 0.30 & 0.01 & 0.30 &$-2.37$ \\               
\object{HD\,128279} & $<-2.01$ & { 0.03} & 0.11 & 0.01 & 0.11 &$<-1.94$\\             
\object{HD\,200654} & $<-2.06$ & { 0.03} & 0.17 & 0.01 & 0.17 &$<-2.04$\\             
\object{HD\,218857} & $-1.42$  & $-0.03$ & 0.22 & 0.03 & 0.22 &$-1.52$\\                  
\object{HD\,274939} & $-1.20$  & $-0.03$ & 0.20 & 0.03 & 0.20 &$-1.28$\\                  
\object{$\mathrm{BD}-01\degr\,2582$} & $-1.74$ & $-0.01$ & 0.25 & 0.03 & 0.25 & $-1.81$\\ 
\object{$\mathrm{CD}-24\degr\,1782$} & $-1.45$ & { 0.00} & 0.21& 0.02 & 0.21 &$-1.54$\\  
\object{$\mathrm{CD}-30\degr\,0298$} & $-2.04$ & { 0.03} & 0.27 & 0.00 & 0.27&$-2.04$\\
\hline
\end{tabular}
\end{center}
\end{table}

The formal uncertainty in our Be abundance determinations was derived
similarly to what was done in the case of lithium. Here
we took only into account the sensitivity of Be to changes (corresponding
to 1$\sigma$ uncertainties) in the stellar parameters. We found that the 
surface gravity determination plays a major role in obtaining accurate Be abundances, leading
to uncertainties $A(\mathrm{Be})$ of about 0.2 dex. On the other hand, 
uncertainties associated to effective temperature and metallicity are 
insignificant.

An important source of error in the Be abundance is due to the normalisation 
and continuum location in fitting the synthetic spectra. We estimate that this
source of error may well contribute errors in $A(\mathrm{Be})$ of about 0.2 dex.
There is also a risk that this error may be systematic, e.g. leading to 
too small Be abundances for the majority of the stars. 

The Be abundances given in Table~\ref{Be} were determined from the LTE spectral 
analysis of the \belines. However, according to the work of {\citet{ramon95}} and 
 Garc\'{\i}a P\'erez and Kiselman\,(2005, in preparation), these lines are formed 
in the stellar atmospheres under non LTE conditions. 
Specific NLTE calculations were carried out following the procedure described
and discussed by Garc\'{\i}a P\'erez and Kiselman\,(2005, in preparation), 
using an adapted version of the {\sc multi} code.
NLTE Be abundances were determined by matching the equivalent width calculated in LTE 
with abundances from the spectrum synthesis with the NLTE computed ones. The highest 
NLTE correction (0.2 dex) is found for the giant star, \object{HD\,126587}. For this star, 
the NLTE Li abundance correction is also among the highest.

\section{Results and discussion}

\begin{table}
\caption{\label{DLiBe} Logarithmic Li and Be depletion factors for the programme stars, 
based on abundances in metal-poor dwarf stars. For Be, factors 
associated to two different Be-vs-O trends are quoted: trend in 
Primas et al.\,(2005, in preparation) (Col.~4) and in \citet{BoesBe99} (Col.~5). Predictions 
based on \citet{Deliyannis90} evolutionary tracks are given in Col.~3
and Col.~6.}
\centering
 \renewcommand{\tabcolsep}{3pt}
\begin{tabular}{lcccccccccc}
\hline\hline 
Star & $D(\mathrm{Li})$ & $D(\mathrm{Li})$ & $D(\mathrm{Be})$ & $D(\mathrm{Be})$ & $D(\mathrm{Be})$ \\
     &  obs  &  theor & obs & Boesgaard  & theor \\
\hline
\object{HD\,4306}   & $-1.14$  & $-1.22$  & $-1.20$ & $-0.83$ & $-0.87$ \\
\object{HD\,26169}  & $-1.12$  & $-1.22$  & $-0.91$ & $-0.51$ & $-0.87$ \\
\object{HD\,27928}  & $-1.08$  & $-1.22$  & $-1.27$ & $-0.87$ & $-0.87$ \\
\object{HD\,45282}  & $-0.96$  & $-1.01$  & $-1.19$ & $-1.01$ & $-0.65$ \\
\object{HD\,108317} & $<-2.03$  & $-1.10$  & $<-1.72$ & $<-1.38$ & $-0.75$ \\
\object{HD\,126587} & $-1.24$  & $-1.22$  & $-0.80$ & $-0.17$ & $-0.87$ \\
\object{HD\,128279} & $-0.96$  & $-1.03$  & $<-1.09$ & $<-0.67$ & $-0.68$ \\
\object{HD\,200654} & $-1.00$  & $-1.11$  & $<-0.95$ & $<-0.46$ & $-0.76$ \\
\object{HD\,218857} & $-1.14$  & $-1.22$  & $-1.09$ & $-0.79$ & $-0.87$ \\
\object{HD\,274939} & $-1.03$  & $-1.22$  & $-1.34$ & $-1.17$ & $-0.87$ \\
\object{$\mathrm{BD}-01\degr\,2582$} & $-1.4$3  & $-1.22$ & $-1.20$ & $-0.84$&$-0.87$ \\
\object{$\mathrm{CD}-24\degr\,1782$} & $-1.04$  & $-1.17$ & $-0.50$ & $-0.02$&$-0.82$ \\
\object{$\mathrm{CD}-30\degr\,0298$} & $-1.12$  & $-1.19$ & $-0.54$ &  0.07& $-0.84$ \\
\hline
\end{tabular}
\end{table}

Simultaneous knowledge of lithium and beryllium abundances in the same
star is a powerful diagnostic of mixing activity in stellar interiors. 
According to standard models, halo stars cooler than about 5700 K 
on the main-sequence should show Li depletion (which has already taken
place during their pre-main-sequence phase). On the contrary, 
metal-poor stars warmer than this temperature all show a similar Li
abundance, the so-called Spite 
lithium plateau \citep{Spite82,Ryan99}. According to stellar 
evolution theory, as stars evolve from dwarfs to giants, they get cooler 
but the surface convection zone deepens and reaches layers which were 
hot enough to ``burn'' Li. The matter poor in Li is mixed with the rest 
of the convection zone, so that dilution of lithium is 
expected.

The surface gravities of our programme stars are typical 
of subgiants. Their spectra show weak lithium lines which suggest
depletion. In one case, \object{HD\,108317}, the Li line was not 
even detected. Something similar happens to the {\belines} in 
subgiants, i.e. they get weaker and more difficult to detect, 
especially the redder line of
the doublet which is the cleaner but unfortunately also the weaker 
of the two. This line ($\lambda$ 313.1 nm) was not detected in the observed 
spectra of \object{HD\,108317}, \object{HD\,128279} and \object{HD\,200654}. 
When upper limits are given, they are constrained by synthetic spectrum fits
to both lines. 

\begin {figure}
\begin{center}
\resizebox{\hsize}{!}
{\includegraphics{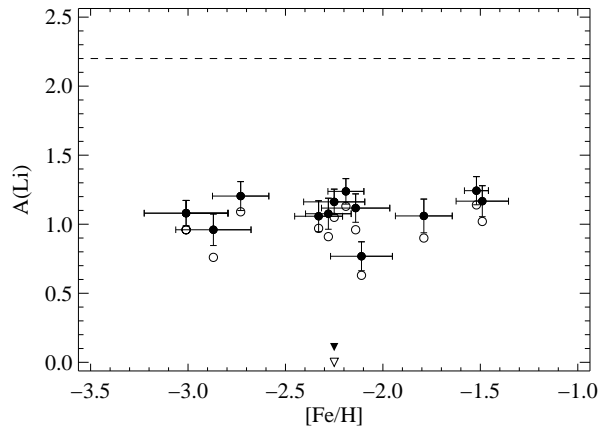}}
\caption{\label{lit} Logarithmic lithium abundances in a 12-scale as a function 
of stellar metallicity. Both NLTE (filled circles) and LTE (open circles) 
abundances are shown. Dashed line shows a 2.2 constant value as representative 
of metal-poor main-sequence stars in the Spite plateau.}
\end{center}
\end{figure}

\begin {figure}
\begin{center}
\resizebox{\hsize}{!}
{\includegraphics{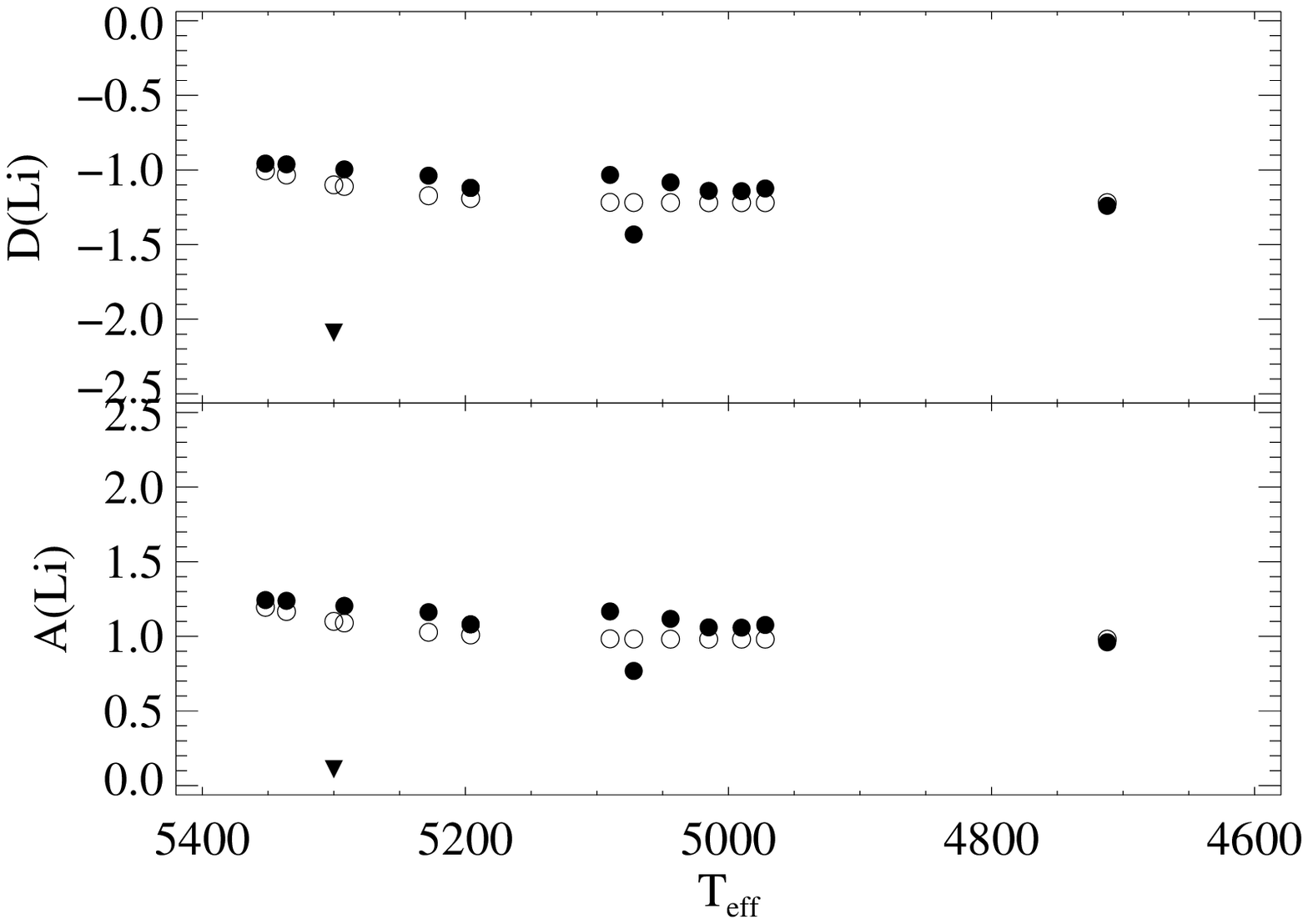}}
\caption{\label{litteff} Logarithmic Li depletion factors (top panel) and 
Li abundances (bottom panel) as a function of effective temperature. Observed 
depletion factors and the NLTE Li abundances (filled circles) on which these 
factors are based are plotted together with the predicted values (open circles).
 A value of 2.2 was taken as representative of initial Li abundance. The triangle 
denotes the star \object{HD\,108317} with an upper limit on the Li abundance.}
\end{center}
\end{figure}

\subsection{Lithium depletion}

As is seen in Figs.~\ref{lit} and \ref{litteff}, the derived lithium
abundances of the programme stars, either the LTE (open circles) or the NLTE
(filled circles) values, are well below the Spite plateau value (here taken
to be $A(\mathrm{Li})_\mathrm{NLTE}=2.20$, close to \citet{Bonifacio97}), 
by about 1 dex. No obvious correlation is seen between the lithium abundances
and metallicity, and the observed difference in $A(\mathrm{Li})$ between stars
at the same metallicity can be fully explained by the associated errors, mostly
due to uncertainties in the effective temperature. 

Assuming that our observed subgiants have left the MS with Li abundance values 
typical of the Spite plateau, we can estimate their logarithmic depletion
factors using the Spite plateau value as the initial value
($A(\mathrm{Li})_{\mathrm{i}}=2.2$). The depletion factors (Col.~2, 
Table~\ref{DLiBe}) based on that MS value are compared with dilution
predictions (Col.~3, Table~\ref{DLiBe}) in Fig.~\ref{litteff} as a
function of effective temperature, where the predicted values are
based on evolutionary tracks presented in \citet{Deliyannis90} for
a subgiant star with $0.775\,M_{\sun}$ and [Fe/H]=$-2.3$. 

First of all, we note that there is now a trend in the logarithmic depletion 
factor $D(\mathrm{Li})$ (where 
$D(\mathrm{Li})=A(\mathrm{Li}) - A(\mathrm{Li})_{i}$) with the
effective temperature. This is intrinsically related to the dredge-up mechanism and other observational 
studies, e.g. \citet{Pilachowski93}, \citet{Lebre99} and \citet{Medeiros00} have
also found a similar depletion pattern. If all the stars have left the main-sequence
with the same Li abundance, then our results imply that the cooler subgiants have
undergone the larger depletions: cooler means more evolved, hence a more complete
dredge-up mechanism, which is in agreement with standard stellar models 
\citep[e.g.][]{Deliyannis90}.

Secondly, the observed depletions of our stars seem to agree rather well with the 
predictions, although the observed stars may systematically show somewhat less
depletion than the models. We note that \object{$\mathrm{BD}-01\degr\,2582$} lies
slightly below the general trend, and that \object{HD\,108317}, denoted by a triangle,
seems to be extremely depleted (especially since its abundance is only an upper limit)
in lithium compared with the rest of the sample. 

Differences between observed and predicted depletion factors seen in
Fig.~\ref{litteff} are of the order of 0.1 dex, which may be
insignificant, since the Spite plateau value is known with an accuracy
hardly better than 0.1 dex. For instance, \citet{Ryan99} suggested a trend
of $A(\mathrm{Li})$ with metallicity, and derived a plateau value of
2.1, whereas \citet{Melendez04} recently found $A(\mathrm{Li})$=2.37, 
using a newly derived (high) temperature scale. This is just to give
an idea of the range of values that one can find in the literature,
but it should be kept in mind that these comparisons are only
qualitative, as no attempt has been made to put these values on
the same scale (i.e. taking into account the fact that different
authors may have used different analytical tools, different model
atmospheres, different temperature scales). 

More relevant to our study which deals with subgiant stars is instead
the very recent analysis by \citet{Charbonnel05}, whose sample includes
not only MS, but also evolved objects. One of their main findings is that
the plateau value for their sample of evolved stars is, on average, slightly
higher (2.235 or 2.259, depending on the lower \teff~cut-off defining the
plateau region, 6000\,K or 5700\,K respectively) than for the dwarfs (2.215
or 2.176, for the same \teff~cutoffs, respectively). Also, the lithium
abundances derived for the evolved stars appear to be slightly more scattered
than those for the MS stars. Although these differences are well within the
observational errors (as quoted by the authors themselves), if real they would
imply a better agreement between observations and predictions for our stars
(i.e. differences could be reduced by up to 0.06 dex, e.g. if we had assumed 
$A(\mathrm{Li})_{i}=2.259$ as our initial lithium abundance).

\begin {figure}
\begin{center}
{\resizebox{\hsize}{!}{\rotatebox{90}{\includegraphics{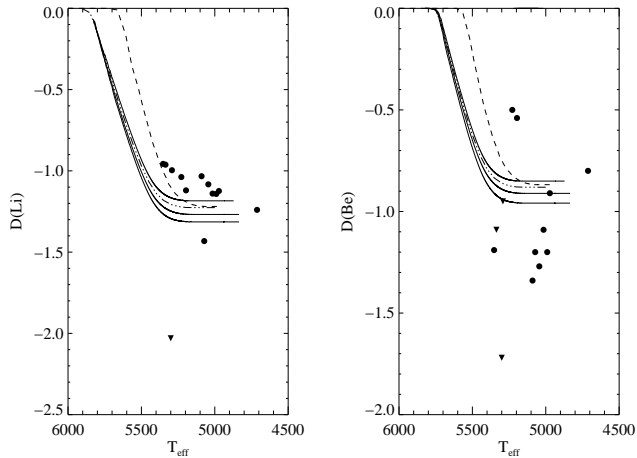}}}}
\caption{\label{charb} Observed logarithmic depletion factors (circles) and predictions based on: 
evolutionary tracks with $\mathrm{[Fe/H]}=-2.27$ and stellar masses of 0.75, 0.80 and 
$0.85\,M_{\sun}$ (solid lines, from top to bottom) from Charbonnel (priv. comm.) and
evolutionary track with $\mathrm{[Fe/H]}=-2.30$ and stellar mass of $0.775\,M_{\sun}$
(dashed line) from Deliyannis. For comparison, an interpolated model from Charbonnel
for a stellar mass of $0.775\,M_{\sun}$ is also presented (dot-dashed line).}
\end{center}
\end{figure}

Clearly, in order to draw firmer conclusions, one should in principle 
compare the observed depletions to more than just one set of theoretical
models. In practice, this is not an easy task, because there have not been
simultaneous theoretical predictions for both Li and Be since 
\citet{Deliyannis90}. For the purpose of this test, Charbonnel ({\it private
communication}) has kindly provided us with three models computed for a
similar stellar metallicity ($\mathrm{[Fe/H]}=-2.27$) of Deliyannis' model but
for three different stellar masses (0.75, 0.80 and $0.85\,M_{\sun}$). As one
can see from Fig.\,\ref{charb}, this new set of models predict different
depletions for different stellar masses. Also, it is interesting to note that
depletion sets in at higher temperatures than in the model of Deliyannis: in
other words, the same amount of depletion is reached at different temperatures
in the two sets of models. This is not completely surprising, as several
factors (e.g. initial He content, mixing length, opacities, equation
of state) may be slightly different in these models.

In order to make as similar comparison as possible to the dilution curves of
\citet{Deliyannis90}, we have interpolated Charbonnel's dilution curves to a stellar
mass of 0.775 $M_\odot$ (dot-dashed curve in Fig.\,\ref{charb}). Inspection
of this figure shows that higher discrepancies are found between observations
and predictions when the interpolated dilution curve is adopted. It seems that for
Charbonnel's dilution curves, masses lower than 0.70 $M_\odot$ would improve the
agreement for the majority of our stars. 

\subsection{Beryllium depletion}

\begin {figure}
\begin{center}
\resizebox{\hsize}{!}
{\includegraphics{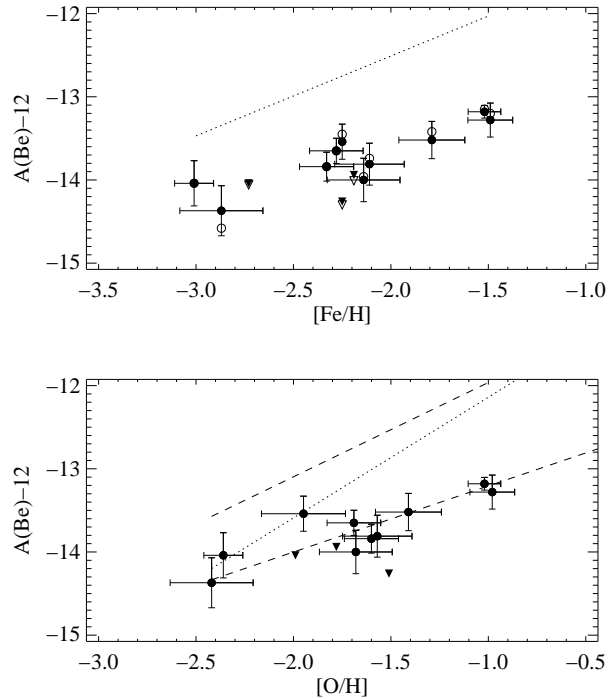}}
\caption{\label{bery} NLTE and LTE Be abundances as a function of 
metallicity (top panel) and only NLTE values as a function of oxygen 
abundances (bottom panel). The dotted lines 
correspond to: the Be-vs-Fe (top panel) and the Be-vs-O (bottom panel) 
correlation for dwarfs from \citet{BoesBe99}. The dashed lines correspond 
to the Be-vs-O correlation for dwarfs from Primas et al.\,(2005, in preparation) 
(upper line in bottom panel) and for our observed subgiants (lower line in 
bottom panel).}
\end{center}
\end{figure}

Our beryllium abundances are plotted versus metallicity in Fig.~\ref{bery}
(top panel), with LTE and NLTE values represented by open and filled circles,
respectively. We note that, as in the case of lithium, all our stars have
beryllium contents much lower than the general Be evolutionary trend derived
from dwarf stars \citep[shown as a dotted line of unitary slope in the figure, 
cf][]{BoesBe99}. However, since the main process responsible for the
formation of beryllium in the Galaxy is cosmic-ray spallation reactions which
involve protons and $\alpha$ particles on one side and C and O nuclei on the
other \citep{Reeves70,Meneguzzi71}, it is more instructive to plot Be versus
oxygen instead of iron.

The bottom panel in Fig.~\ref{bery} shows the correlation between NLTE
beryllium and oxygen abundances. The oxygen abundances were derived from
the forbidden oxygen line at 630.03 nm \citep{aegpO}. The best fit we
obtained is a linear fit (dashed line, with slope of 0.8). Stars with
only upper limits (denoted by triangles) on their Be abundances were
excluded from the fit, as well as \object{CD-24\degr\,1782} (its oxygen
abundance comes from OH UV lines since its forbidden line was not detected). 

As was also the case for lithium, \object{HD\,108317} seems to stick out
from the rest of the sample, with a beryllium content lower than other
stars with similar parameters and oxygen abundances. The observed pattern
could imply that this star has undergone a more severe depletion or it may
be that its Be abundance before reaching the subgiant evolutionary phase 
was lower than for stars of similar oxygen abundances, due to different
production scenarios. On the contrary, in the case of
\object{$\mathrm{BD}-01\degr\,2582$}, the beryllium content is comparable 
to other stars of the sample with similar oxygen abundances, despite its 
lithium being lower than our derived trend. This is not remarkable, since 
lithium burns at a lower temperature, i.e. earlier than beryllium if the 
mixing deepens to hotter stellar layers.

\begin{figure}
\begin{center}
\resizebox{\hsize}{!}
{\includegraphics{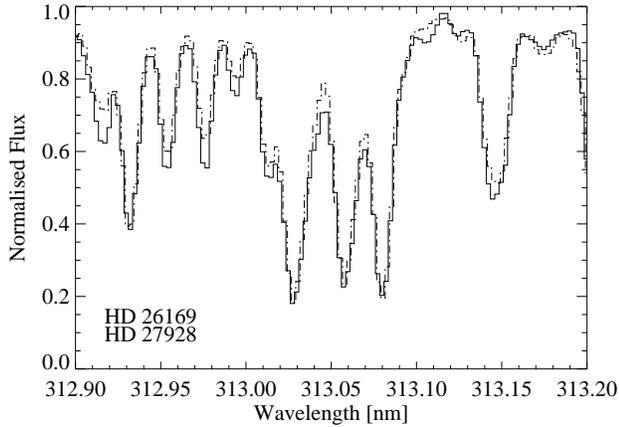}}
\caption{\label{comparison} The UV \ion{Be}{ii} doublet lines in the observed
spectra of \object{HD\,26169} (solid line) and \object{HD\,27928} (dot-dashed line).}
\end{center}
\end{figure}

Furthermore, the stars \object{HD\,26169} and \object{HD\,27928} have very
similar O abundances ($\mathrm{[O/H]}\simeq-1.7$), but their beryllium
contents differ by 0.35 dex. We note that this difference can be only
marginally accounted for by the associated observational errors (the final
error in our Be abundances is $\simeq0.20~\mathrm{dex}$). As it can be seen
from Fig.~\ref{comparison}, where the spectra of the two stars are compared,
the difference in Be abundances may well be real. Having already measured
lithium in both stars, we can now use this important information to check if
there is any evidence for different mixing history in these two stars. From
Table~\ref{Li}, we see that \object{HD\,26169} and \object{HD\,27928} have
practically identical lithium abundances (within 0.05 dex from each other)
and very similar fundamental stellar parameters (cf Table~\ref{gripar}). One 
could then expect that they have also undergone similar mixing processes: 
if Be is affected by a given transport mechanism, then the Li should have 
changed at least by the same amount. Therefore, the only possible explanation, 
if the Be abundances are proven to depart significantly in these two stars, is 
a difference in their initial Be abundances: a production more than a des\-truc\-tion 
effect.

In order to take a look at the global trend of our Be abundances, we now
compare the Be-vs-O correlation we have derived from our subgiant sample
with two linear correlations (dotted and upper dashed line, bottom panel,
Fig.~\ref{bery}) derived for dwarfs respectively by \citet{BoesBe99} and
more recently also by our group (Primas et al.\,2005, in preparation).
We note that the latter analysis finds a non-negligible scatter around
the high-metallicity end of the Be-vs-O relation (which overlaps with our
more metal-rich stars) and that both linear and broken fits are under
investigation. However, for the purpose of our discussion, and because
the data of Primas et al. are still preliminary, we chose to use the
simplest of their fits, i.e. the linear one. 

Under the assumption that the initial Be abundance of all our subgiants
was the one found in dwarf stars, we can now estimate logarithmic Be
depletion factors ($D(\mathrm{Be})$), based on the NLTE Be abundances
reported in Table~\ref{Be}. It is important to note that because both
dwarf correlations reach only [O/H]=$-$2.0 (whereas our sample reaches
[O/H]$\simeq-$2.5), we decided to extrapolate both fits down to our
lowest O abundance in order to be able to compute the D(Be) factors for 
the entire subgiant sample. The factors thus derived suggest that our stars
have depleted Be by approximately 1 dex, as they have evolved off of the
main sequence.

\begin {figure}
\begin{center}
\resizebox{\hsize}{!}
{\includegraphics{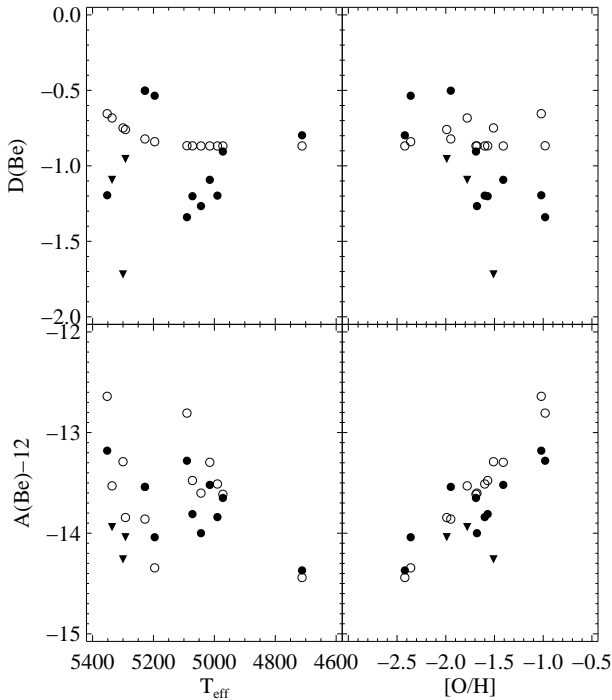}}
\caption{\label{beryteff} Logarithmic Be depletion factors (top panels) and Be abundances 
(bottom panels) as a function of effective temperature (left panels) and oxygen 
abundance (right panels). Our estimates 
(filled circles) are compared with the predictions (open circles). Stars 
with upper limits on Be abundance are denoted by triangles.}
\end{center}
\end{figure}

Similarly to Li, we compare our estimates (Col.~4, Table~\ref{DLiBe}) with 
predictions (Col.~6, Table~\ref{DLiBe}) based on depletion calculations 
kindly provided by C.~Deliyannis (private communication). As it is shown in 
the upper panel of Fig.~\ref{beryteff}, the standard models predict
a Be dilution in subgiants which is expected to depend on effective
temperature. However, our depletion factors do not show any obvious sign
of such a dependence, although the observational or cosmic scatter may be
too large to disclose such a tendency. 

Because of the different slopes in the dwarf correlations we have used as
a comparison, the emerging trends are quite different. Be seems to be
systematically more depleted in our subgiant sample (on average by 0.2 dex)
than expected from the depletion calculations (compare filled with empty
circles) if the Primas et al. correlation is used. Instead, when the
\citet{BoesBe99} relation is considered, the difference between observed and
predicted depletion values is reduced (Cols.~5 and 6, Table~\ref{DLiBe}),
and for some objects we even observe a reversed behavior, i.e. the models
seem to overproduce the amount of depletion that Be has already undergone.
Note, however, that the relation of \citet{BoesBe99} is based on oxygen
abundances derived from OH UV lines instead of the forbidden oxygen line
at 630.03 nm. For these reasons and because we and Primas et al. have used
similar methods to derive abundances and stellar parameters, we believe that
it is more appropriate to carry out the comparison between MS and subgiant
stars in terms of the Primas et al. dwarf relation.

Furthermore, the derived depletion factors do not follow a clean trend but
they are quite scattered, especially in a range of 150 K around 5050 K, where
the spread in the Be depletion factors is considerable (with a maximum
difference of the order of 0.4 dex). Although this seems to be at odds with
the constant depletion value expected from dilution at these temperatures,
the observational errors associated to our Be abundances are quite large.
Therefore, the observed scatter cannot be taken as a clear indication of
physical differences between the stars. 

Possibly, the systematic discrepancies between observed and predicted
depletion factors for Be may be the product of our assumption that initial
Be abundances are those found in dwarf stars. We explore this possibility
by comparing estimates with predictions in a $D(\mathrm{Be})$-vs-[O/H] plot
(bottom panel in Fig.~\ref{beryteff}). We do not see, however, significant
differences in depletion discrepancies for different [O/H]. This suggests
that if the assumed values for the initial Be abundance were wrong, the
relative errors would be independent of [O/H]. 

According to the tracks of Charbonnel and as it was the case for Li, Be 
depletion depends on the stellar mass. It is also the case here that the 
Charbonnel's dilution curve interpolated for a stellar mass of $0.775\,M_{\sun}$ 
produces more depletion than the dilution curve of Deliyannis (see right panel in 
Fig.\ref{charb}). This clearly helps to reduce the differences between 
observations and predictions, although this is not enough to eliminate them completely. 
For most stars in our sample and unlike the case of Li, it would be necessary 
to resort to dilution curves corresponding to stellar masses higher than 
$0.775\,M_{\sun}$, in order to reconcile our results with the predictions. 
Furthermore, we note that differences between observations and predictions 
are much higher than the differences between the dilution curves shown here.

 \begin {figure}
\begin{center}
\resizebox{\hsize}{!}
{\includegraphics{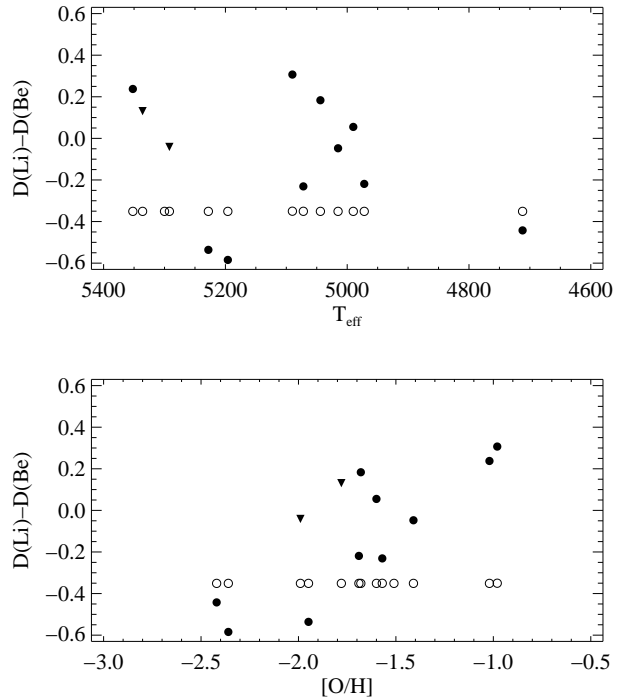}}
\caption{\label{litberteff} The differences between logarithmic depletion factors for Li and 
Be as a function of temperature (top panel) and oxygen abundance (bottom panel). Estimates 
(filled circles) are compared with predictions based on \citet{Deliyannis90} 
(open circles).}
\end{center}
\end{figure}

\subsection{Comparison between Li and Be depletions}

Lithium is the most fragile of the three light elements. If the Be 
content of a star is depleted, it is expected to have already depleted 
all, or a great fraction of its lithium. 

According to standard model predictions (open circles, Fig.~\ref{litberteff}),
Li depletion should be around 0.3-0.4 dex higher than Be depletion.
Despite this is a feature common to both sets of dilution curves we have used
in our comparisons, most of our observed subgiants (more than 50\%) are
characterised by much larger Be depletions (compared to their amount of Li
depletion). The stars with a detected Li line but with only upper limits
for the Be abundance (\object{HD\,128279} and \object{HD\,200654}) are of
special interest in this respect.

As already discussed in Sect.~5.1, the agreement between observed and
predicted D(Li) values is at the level of 0.1--0.15 dex, and it holds for
both sets of dilution curves (i.e. Deliyannis and Charbonnel) in the case of the
cooler objects of our sample. Thus, it must be beryllium that sticks out:  
models predict a much smaller amount of depletion, between 0.25 and 0.65 dex,
depending on the star under consideration and also on the set of models (for
Be, Charbonnel's dilution curves predict a larger amount of depletion for the warmer
stars of our sample). Furthermore, one feature that could be worth
further investigating is the metallicity dependence of the theoretical
dilution curves: in fact, we noted that the best agreement between observed and
predicted depletion factors (when both Li and Be are taken into account
simultaneously) is obtained for the two most metal-poor stars of our sample
(\object{$\mathrm{CD}-24\degr\,1782$} and \object{$\mathrm{CD}-30\degr\,0298$}), 
while the worse agreement is obtained for the most metal-rich stars of our sample 
(\object{HD\,45282} and \object{HD\,274939}).

Of course, another possibility is that our Be abundances 
are systematically too low by about 0.2 dex as a consequence of a too low 
spectrum normalisation (continuum location). This may be directly seen in 
Fig.~\ref{syn}, where we would not need to change the continuum level very 
much in order to get a good fit of the observed \ion{Be}{ii} lines with the 
lower solid lines. With such Be abundance uncertainty in mind, we should 
refrain from any definitive statement on departures from the standard 
depletion calculations, as well as on the possibility that the Be depletion
could be larger than that of Li. However, we can make some qualitative
speculations.

If our abundances are taken at face value (i.e. without any systematic errors
in the Be abundances), we could still reconcile the Li and Be depletion data by
advocating a higher Li depletion, e.g. assuming a higher initial lithium of 
$A({\mathrm{Li}})=2.6$. As a matter of fact, the primordial lithium abundance inferred by the
baryonic density obtained from WMAP data \citep{Spergel03} in combination
with standard Big Bang nucleo--synthesis \citep[SBBN,][]{Coc04} is 
$A({\mathrm{Li}})=2.6$. However, since much lower values are consistently observed in
metal-poor dwarf stars (and most of them cluster around 
$A({\mathrm{Li}})=2.2$), such a scenario is unlikely. Mixing models like those including rotationally 
induced mixing \citep{Pinsonneault02} indeed suggest some degree of depletion 
in these stars with respect to the primordial value, the magnitude of which strongly depends on stellar properties 
such as rotational velocities, masses, temperatures, metallicities etc. 
However, the absence of any significant scatter in the Li abundances measured
in warm halo stars does not favour such a scenario \citep{Bonifacio97,Ryan99},
and it is not obvious whether mixing processes are able to produce such high
Li and Be depletions. We note that theo\-re\-ti\-cal computations including rotation mixing 
and gravity waves seem to be able to deplete Li in a very homogeneous way \citep{Talon04}, as 
required by what is observed in warm halo dwarf stars. Unfortunately, Be depletions have not yet 
been derived and inspected. The luminosities of our metal-poor subgiants suggest that
several of them are starting to ascend the red giant branch so they may already
have suffered a first dredge-up. If this process is not well described by
standard theory, the predictions of both Li and Be may be underestimated.

\section{Conclusions}

As stars evolve off the main sequence, their surface convection zone starts to 
move inwards reaching first Li depleted and later Be depleted matter. The mixing
between these outer and inner layers reduces the surface abundances of these
light elements. Subgiants are expected to show signs of such depletion in their
Li and Be abundances as compared with MS values. Our derived Li abundances lie
significantly below the Spite lithium plateau. The depletion is approximately
one order of magnitude. A trend with effective temperature may be traced: the
warmer stars show the higher Li abundances. In general, these abundances agree
well with the standard predictions from evolutionary models of
\citet{Deliyannis90} with the provision that the observations may show somewhat
less Li depletion than expected. These differences can be reduced to zero if a
value for the Spite plateau 0.1 dex higher were assumed instead, which is quite
possible given the uncertainty of its value.

As far as Be is concerned, observations of subgiants show that abundances of
Be and O are correlated, as in the case of dwarf stars. Be depletion estimates
depend on the assumed Be-vs-O trend of dwarfs, e.g. in our case, that of
Boesgaard et al. (1999) and of Primas et al. (2005, in preparation). However,
since similar stellar parameters and analytical methods were used here and by
Primas et al., we have considered the latter source more appropriate for
estimating our Be depletion factors. What emerges is that observed and
predicted depletions in our subgiant sample are comparable but in some
conflict; the predicted Be depletion is typically 0.3 - 0.4 dex smaller than 
what we observe. We note that a higher initial amount of Li would help 
reconciling our derived $D(\mathrm{Li})$ and $D(\mathrm{Be})$ depletion factors, 
so that $D(\mathrm{Li})$ is larger than $D(\mathrm{Be})$ for all our stars. This 
is challenged by the robustness of the Spite's plateau value of 
$A(\mathrm{Li})\simeq2.2$. A wrong temperature scale could 
affect this value, but not enough to reconcile our $D(\mathrm{Li})$ and $D(\mathrm{Be})$ values. 
This would require $A(\mathrm{Li})\sim2.6$ which is an unlikely 
value for the initial Li content of dwarf stars but surprisingly matches 
the WMAP$+$SBBN predictions for the primordial Li content. An alternative would be depletion
mechanisms acting more efficiently for Be than for Li, mechanisms that are
not easy to envisage. However, we note that 0.3 dex could also be
a (conservative) estimate of the uncertainty in our Be abundances due to
the continuum placement. Hence, should we have systematically performed our
spectrum normalisation too low, then a better agreement between standard
predictions and observations is obviously obtained.

Similar conclusions can be derived when Charbonnel's dilution curves 
are used: namely relatively good agreement between observations and predictions
of Li depletion, but somewhat worse in the case of beryllium. According to
these dilution curves, our results on Li suggest that the observed
subgiants should have masses smaller than $0.75\,M_{\sun}$, whereas our
observed Be depletion factors suggest that a better agreement could be
achieved for stellar masses larger than $0.85\,M_{\sun}$.

Further disentangling between light element observed abundances and
theoretical predictions require the availability of a grid of dilution curves,
covering the whole range of metallicities of our sample and different masses
(since there seem to be a dependence on the stellar mass), as well as higher
quality data, in order to reduce the error-bars associated to individual
abundance measurements. Should our Be abundances and their depletion factors
be confirmed, then this will have important implications on the modelling
of the interiors of these stars.

\begin{acknowledgements} The authors are indebted to C.~Deliyannis 
and C.~Charbonnel for kindly providing us their Li and Be depletion
isochrones. A very important part of the work presented here was carried
out during the visit of A.E.G.P. to ESO and she is very thankful to ESO
for hospitality and financial support. M.~Asplund, B.~Gustafsson, and
C.~Charbonnel have contributed significantly to this work with fruitful
discussions and valuable suggestions. The work has been supported by the 
Nordic Optical Telescope and the Swedish Research Council.

\end{acknowledgements}

\bibliographystyle{aa}
\bibliography{ref}
\end{document}